\documentclass[
aps,
twocolumn,
pra,
longbibliography,
groupedaddress,
superscriptaddress,
showpacs,
letterpaper,
amsfonts,
floatfix,
10pt
]
{revtex4-1}
\RequirePackage{amsmath,amssymb,bm,wasysym}
\RequirePackage{graphicx}
\RequirePackage{hyperref}
\hypersetup{
breaklinks={true},
citecolor={blue},
colorlinks={true},
linkcolor={blue},
pdfauthor={\textcopyright Matthias Droth},
pdfcreator={\LaTeX and\flqq hyperref\frqq},
pdffitwindow={true},
pdfmenubar={true},
pdfpagelayout={OneColumn},
pdfstartview={FitH},
pdftoolbar={true},
plainpages={false},
pdfpagemode={UseThumbs},
}
\begin{document}
\pagestyle{plain}
\title{Electron-electron attraction in an engineered
electromechanical system}
\author{G\'{a}bor Sz\'{e}chenyi}
\affiliation{Institute of Physics, E\"{o}tv\"{o}s University, 1518 Budapest, Hungary}
\author{Andr\'{a}s P\'{a}lyi}
\affiliation{Institute of Physics, E\"{o}tv\"{o}s University, 1518 Budapest, Hungary}
\affiliation{Department of Physics, Budapest University of Technology and Economics, 1111 Budapest, Hungary}
\affiliation{MTA-BME Condensed Matter Research Group, 
Budapest University of Technology and Economics, 1111 Budapest, Hungary}
\author{Matthias Droth}
\thanks{Corresponding author: matthias.droth@mail.bme.hu}
\affiliation{Department of Physics, Budapest University of Technology and Economics, 1111 Budapest, Hungary}
\begin{abstract}
Two electrons in a quantum dot repel each other: their 
interaction can be characterized by a positive interaction energy. 
From the theory of superconductivity, we also know that 
mechanical vibrations of the crystal lattice can make the 
electron-electron interaction attractive. 
Analogously, if a quantum dot interacts with a mechanical degree of 
freedom,  the effective interaction energy can be negative; 
that is, the electron-electron interaction might be attractive. 
In this work, we propose and theoretically study an engineered 
electromechanical system that exhibits 
electron-electron attraction: 
a quantum dot suspended on a nonlinear mechanical
resonator, 
tuned by a bottom and a top gate electrode. 
We focus on the example
of a  dot embedded in a suspended graphene ribbon, 
for which we identify conditions for electron-electron attraction.
Our results suggest the possibility of electronic transport via
tunneling of packets of multiple electrons in such devices, 
similar to that in superconducting nanostructures, 
but without the use of any superconducting elements. 
\end{abstract}
\maketitle
\section{Introduction}
Two electrons usually repel each other due to 
the Coulomb force. 
However,
mechanical vibrations of a crystal lattice can mediate 
an effective attractive interaction between the
delocalized electrons, 
leading to the formation of Cooper pairs and 
the emergence of superconductivity\cite{Onnes1911,Bardeen1957}. 
Moreover, recent experiments have demonstrated 
attractive interaction in the absence of superconductivity,
between electrons confined in
engineered nanostructures:
in a carbon nanotube double quantum dot\cite{Hamo2016},
where the attraction was induced by capacitive coupling
to a nearby auxiliary quantum-dot system\cite{Little1964}, 
and in a sketched quantum dot at the
SrTiO$_3$/LaAlO$_3$ interface\cite{Cheng2015,Cheng2016},
where the mechanism of attraction has not been revealed.
A recent proposal\cite{Butler2015} describes how to engineer 
electron-electron attraction in an artificial 
nanostructure using a careful design
of orbital and tunneling energies. 

\begin{figure*}
\centering
\includegraphics[width=\linewidth]{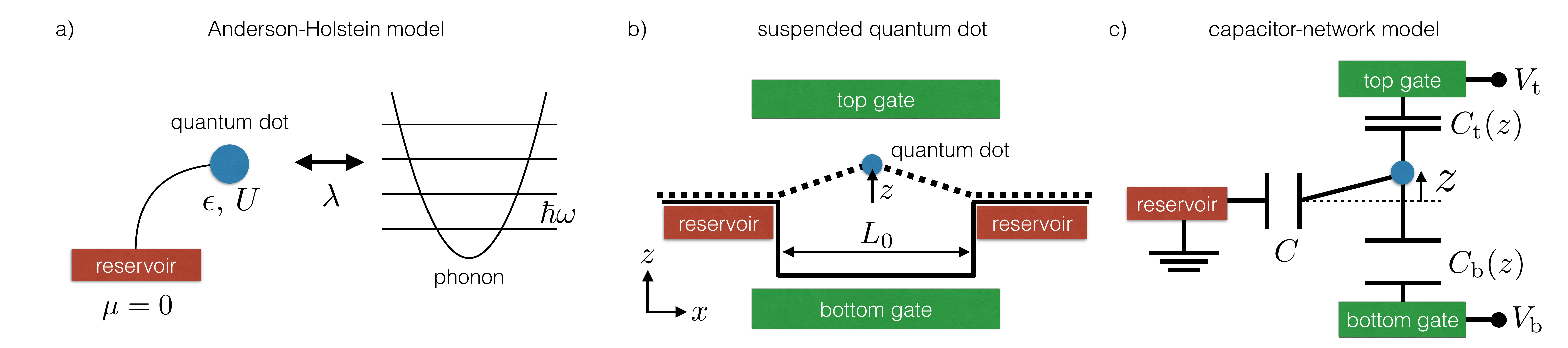}
\caption{
\textbf{
Electromechanical systems showing mechanically assisted
electron-electron attraction.}
(a) Electron-electron attraction arises in the Anderson-Holstein model,
where the charge on a single electronic orbital (quantum dot) 
interacts with a single vibrational mode (phonon). 
(b) Electron-electron attraction can also 
be engineered in a suspended quantum dot (blue spot), which
is located on a nonlinear nanomechanical 
resonator (dashed line).
The dot can be displaced along the $z$ direction.
The equilibrium occupation and displacement of the
dot can be tuned by the top and bottom gate voltages,
$V_{\rm t}$ and $V_{\rm b}$, respectively. 
(c) Capacitor-network model of the suspended dot,
which is coupled to a grounded charge reservoir 
via capacitance $C$,
and to top and bottom gates via $z$-dependent gate capacitances 
$C_\text{t}(z)$ and $C_\text{b}(z)$.
\label{fig:setup}}
\end{figure*}

The possibility of vibration-mediated 
attractive interaction among confined electrons 
has been discussed, e.g.,
in the context of 
amorphous semiconductors\cite{Anderson1975},
vacancies in silicon\cite{Baraff_prl,Baraff_prb},
fullerenes\cite{Lannoo},
and molecular junctions\cite{Koch2006,Ojanen,Koch_chargekondo,Hwang_pairtunneling}.
Figure \ref{fig:setup}a depicts the 
simplest model capturing the basic ingredients of the effect:
it 
involves 
(i) a single vibrational mode (phonon),
characterized by a mass $m$ and 
frequency $\omega$, 
(ii) a single electronic orbital that can be occupied by
one or two electrons, 
i.e., the occupation number is $N \in \{0,1,2\}$;
this orbital is characterized by an on-site
energy $\epsilon$ and 
 a repulsive Coulomb energy $U>0$,
(iii) the coupling between the phonon 
and the confined charge, 
characterized by the force $\lambda$
encoding the coupling strength,
and
(iv) a (zero-temperature) electron reservoir,
with Fermi energy $\mu = 0$, 
which can supply electrons to the orbital.
We refer to this
as the Anderson-Holstein model\cite{Hwang_pairtunneling}.
(See Methods for more details.)
The presence of the electron-phonon coupling
leads to an effective Coulomb energy
$U_\text{eff} = U - \frac{\lambda^2}{m\omega^2}$,
which becomes negative if the coupling $\lambda$
is strong enough, i.e., if $\lambda > \omega \sqrt{mU} $.
That is, a strong enough electron-phonon coupling implies
an attractive electron-electron interaction.

In a system with tunable on-site energy and tunable 
electron-phonon coupling strength, this attractive
electron-electron interaction could lead to remarkable equilibrium
and transport properties, as indicated in Fig.~\ref{fig:results}a and b.
Figure \ref{fig:results}a shows the
\emph{charge stability diagram}, that is, the number  $N_\text{eq}$
of electrons occupying the orbital in equilibrium at zero temperature, as
the function of the two tunable parameters $\epsilon$ and $\lambda$.
For weak electron-phonon coupling
$\lambda <  \omega \sqrt{mU}$, the filling sequence of the 
orbital is regular: 
for example, at $\lambda = 0$, 
as $\epsilon$ is decreased, the
occupation of the orbital increases by one at $\epsilon =0$
and again by one at $\epsilon = -U$. 
In contrast, for strong electron-phonon coupling
$\lambda > \omega \sqrt{mU}$, the occupation
is increased abruptly 
by two as the 0/2 boundary, i.e., 
the boundary between the $N_\text{eq}=0$ and
$N_\text{eq}=2$ regions, is crossed.
When tuned to the 0/2 boundary, such a system is expected
to show an exotic transport effect when embedded between
a source and a drain electrode, reminiscent of 
Cooper-pair transport in a normal-superconductor 
junction\cite{Jehl}:
current is carried by tunneling of 
electron pairs\cite{Koch2006,Hwang_pairtunneling}.
Furthermore, 
 the current -- bias voltage  curve exhibits
a smooth `Coulomb hill' instead of a sharp Coulomb 
plateau\cite{Koch2006};
shot noise\cite{Hwang_pairtunneling} 
(Fano factor) increases compared
to its value for single-electron  
tunneling, corresponding to an increased granularity of 
the charge quanta carrying the current;
and unconventional Coulomb-blockade features are induced
also in the regimes of single-electron tunneling\cite{Fang_bipolaronic}.

The steady progress in the fabrication of molecular junctions
 allows for electrical control of the orbital energies
\cite{HongkunPark,Kubatkin};
however, tuning the strength of the electron-phonon coupling
in these systems is very challenging. 
Accordingly, 
to our knowledge, the effects discussed above have not been
observed in molecular junctions.

In this work, we propose an engineered nanostructure 
to observe electron-electron attraction, 
electron-pair tunneling, and the associated 
interesting phenomenology discussed above. 
The structure we suggest is a 
suspended quantum 
dot\cite{Steele_science,Lassagne_science,Benyamini2014,GangLuo},
see Fig.~\ref{fig:setup}b, 
with a top and a bottom gate electrode. 
This combination allows for
independent 
control of the orbital energy and the electron-phonon coupling strength:
in short, the average gate voltage defines the former, whereas
the gate-voltage difference defines the latter. 

Importantly, in this setup the electron-phonon coupling is
of extrinsic origin\cite{Steele_science,Lassagne_science,Benyamini2014,GangLuo}, 
i.e., it arises due to the external electric field
created by the gates, and not due to intrinsic mechanisms 
(e.g., deformation potential, bond-length change).
Utilizing this extrinsic mechanism brings two 
advantages: 
the electron-phonon coupling is tunable via
the gate voltages, and the corresponding 
extrinsic force can 
well exceed those arising from the intrinsic mechanisms
(see Methods). 
We focus on the example
of a  dot embedded in a suspended graphene ribbon\cite{GangLuo}, 
for which we identify conditions for electron-electron attraction.
Furthermore, our results reveal the possibility of electronic transport via
tunneling of packets of multiple electrons in such devices, 
similar to that in superconducting nanostructures, 
but without the use of any superconducting elements. 
\section{Results}

\begin{figure}
\centering
\includegraphics[width=\columnwidth]{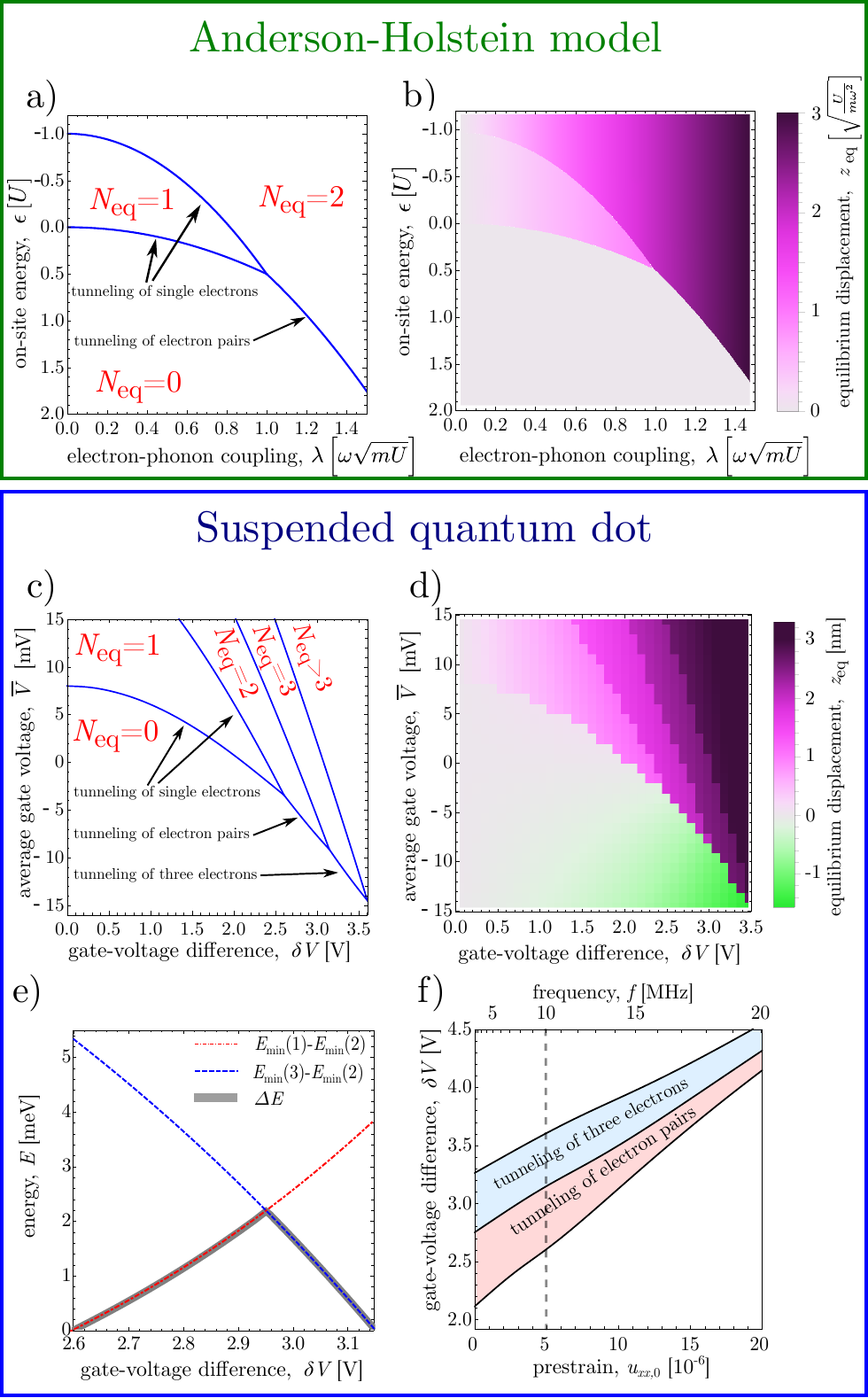}
\caption{(Color online) 
\textbf{Characteristics of mechanically 
assisted electron-electron attraction.}
Charge stability diagram (a) and displacement stability
diagram (b) of the Anderson-Holstein model.
Analogous results for the 
suspended quantum dot are shown in (c) and (d), 
respectively \cite{footnote}. 
(e) Charge excitation gap $\Delta E$ (thick gray line)
along the 0/2 boundary
shown in (c) (labelled as `tunneling of electron pairs').
(f) Gate-voltage range of the 0/2 (pink) and 0/3 (blue) boundaries,
as a function of prestrain.
The upper horizontal axis corresponds to the 
frequency $f$ associated to the resonator (see Methods).
Parameters: ribbon length $L_0=1.5\, \mu\text{m}$,
ribbon width $W = 0.4\, \mu \text{m}$,
ribbon-gate distance $d=150 \, \text{nm}$,
capacitance $C = 5 \,  \text{aF}$.
The prestrain in (c)-(e) is $u_{xx,0} = 5 \times 10^{-6}$.
\label{fig:results}}
\end{figure}

\emph{Setup.}
We consider a quantum dot embedded in a
mechanical resonator, as shown in Fig.~\ref{fig:setup}b.
For concreteness, we formulate a model for the case when the
resonator is a graphene nanoribbon, suspended over
a trench \cite{Bunch2007,Eichler2011,Allen2012,Chen2016}, as
shown in Fig.~\ref{fig:setup}b.
The system is controlled by voltages $V_\text{t}$ and 
$V_\text{b}$ applied on the top and bottom gate electrodes, 
respectively. 
The geometry of the resonator is characterized by 
the width $W$ and the length $L_0$ of the suspended 
part of the ribbon.
The ribbon might be stretched even if the
gates are inactive; characterized by the residual strain
(prestrain) $u_{xx,0} = (L_0-L_u)/L_u$, where 
$L_u$ is the unstretched length of the suspended part of the ribbon.

The dot, indicated by the blue spot in Fig.~\ref{fig:setup}b,
is located on the suspended part of the resonator.
The dot interacts with an electron reservoir;
the capacitive part of this
interaction is characterized by the capacitance $C$.
In addition, electrons can also tunnel between the reservoir and the dot.
The dot is coupled to the top and bottom gates via the
displacement-dependent capacitances $C_\text{t}(z)$ and
$C_\text{b}(z)$, respectively.
As indicated in Fig.~\ref{fig:setup}c, the displacement dependence 
of the capacitances arises since the displacement of the
dot changes the distance between the capacitor plates.
To keep the number of parameters to the minimum, we
assume that the three capacitances are equal at
$z=0$, and that the displacement dependencies are
that of a planar capacitor 
$C_\text{t}(z) = C_\text{b}(-z) = C d/ (d-z)$,
where $d$ is the distance between the 
plates. 
First, we consider the case of an `n-type semiconducting' dot, 
by which we mean that the number $N$ of excess electrons
in the dot at zero gate voltages is zero,
and at finite gate voltages it can only be non-negative, $N \geq 0$. 

\emph{Charge stability diagram.}
Our primary goal is to determine the charge
stability diagram of the dot; that is, 
to determine how the
number $N_\text{eq}$ of electrons in the dot at
zero-temperature equilibrium
depends on the gate voltages $V_\text{t}$ and $V_\text{b}$. 
Motivated by the result of the Anderson-Holstein model
(Fig.~\ref{fig:results}a), 
we look for the 0/2 boundary that separates
regions of the empty dot ($N_\text{eq} = 0$) and the doubly 
occupied dot ($N_\text{eq} = 2$).
To this end, we express the total energy
$E(N,z)$
of the system (see Methods)
which depends on the parameters
$W$, $L_0$, $u_{xx,0}$, $C$, $d$, $V_\text{t}$
and $V_\text{b}$,
as well as the dot occupation $N$ and the 
dot displacement $z$.
We take into account the geometrical nonlinearity of
the resonator by keeping the mechanical energy term
that is of fourth order in the displacement $z$;
this is required to avoid the apparent charge
instability arising in the case of a purely harmonic 
oscillator\cite{Ojanen}.
Then, we minimize the total energy $E(N,z)$  
with respect to $N$ and $z$,
to obtain the zero-temperature equilibrium 
occupation $N_\text{eq}$ and displacement $z_\text{eq}$.

The charge stability diagram 
is shown in Fig.~\ref{fig:results}c, 
for a certain realistic parameter set (see caption).
To be able to compare the diagram with that of the Anderson-Holstein
model (Fig.~\ref{fig:results}a),
we plot $N_\text{eq}$ as the function of the average
gate voltage $\bar{V}=(V_\text{t} + V_\text{b})/2$ and the 
gate-voltage difference $\delta V =(V_\text{t} - V_\text{b})/2$:
intuitively, $\bar V$ controls the on-site energy of the
dot, whereas $\delta V$ controls the electric field that acts on the
dot and hence controls the coupling strength between the dot charge
and the resonator.

The key features in Fig.~\ref{fig:results}c are as follows. 
(i) Overall, the diagram shows strong qualitative similarities
with that of the Anderson-Holstein model (Fig.~\ref{fig:results}a). 
(ii) Similarly to the Anderson-Holstein case (Fig.~\ref{fig:results}a),
Fig.~\ref{fig:results}c also shows a triple point between
the $N_\text{eq} = 0,1,2$ regions.
The coordinates of this triple point are
$\delta V_2 \approx 2.6 \, \text{V}$ and
$\bar{V}_2 \approx  - 3.5\, \text{mV}$.
(iii) In addition to the Anderson-Holstein result, the figure 
also shows a
triple point between the $N_\text{eq} = 0,2,3$ 
regions, at 
$\delta V_3 \approx 3.1 \, \text{V}$ 
and
$\bar{V}_3 \approx \, -9 \, \text{mV}$, 
and a triple point between the $N_\text{eq} = 0,3,4$
regions, 
at $\delta V_4 \approx 3.6 \, \text{V}$ and
$\bar{V}_4 \approx -14.5 \, \text{mV}$.
(iv) Similarly to the Anderson-Holstein case,
a 0/2 boundary is observed in Fig.~\ref{fig:results}c,
labelled as `tunneling of electron pairs'.
This 0/2 boundary connects the
triple points at 
$(\delta V_2,\bar{V}_2)$ and
$(\delta V_3,\bar{V}_3)$.
(v)  Importantly, 
Fig.~\ref{fig:results}c shows that the 0/2 boundary 
arises at a gate-voltage difference of a few volts, 
and an average gate voltage of a few millivolts, 
which suggests that the experimental observation 
of this charge stability diagram, and the 
pair-tunneling transport effects it implies\cite{Koch2006}, 
is feasible. 
(vi) In addition to the 0/2 boundary, 
Fig.~\ref{fig:results}c also shows 
0/$N_\text{eq}$ boundaries with 
$N_\text{eq} > 2$; e.g., 
the 0/3 boundary
between the triple points $(\delta V_3,\bar{V}_3)$
and $(\delta V_4,\bar{V}_4)$,
labelled as `tunneling of three electrons'. 

\emph{Displacement stability diagram.}
Besides the charge stability diagram, it is instructive,
and, for the description of transport effects, is crucial to 
describe how the equilibrium 
displacement $z_\text{eq}$
varies as the gate voltages are tuned. 
We will refer to this function as the 
\emph{displacement stability diagram}.
Figure \ref{fig:results}b shows the displacement 
stability diagram of the Anderson-Holstein model, 
whereas Fig.~\ref{fig:results}d shows
the diagram for our suspended quantum dot model,
for a specific set of parameter values listed in the caption. 
(The pixel structure in Fig.~\ref{fig:results}d is due to data
being obtained on a $\delta V$-$\bar{V}$
grid of size $36 \times 30$.)
Naturally, the equilibrium displacement varies smoothly
within the regions belonging to a certain $N_\text{eq}$,
and jumps abruptly along the boundaries between 
those regions.
Note that 
for the suspended quantum dot, the characteristic scale of these
jumps is nanometer. 

The Anderson-Holstein displacement stability diagram
(Fig.~\ref{fig:results}b)
shows zero displacement $z_\text{eq} = 0$ for an uncharged dot, 
i.e., in the $N_\text{eq} = 0$ region. 
However, the uncharged 
suspended quantum dot can be displaced
toward the bottom gate, as indicated by the green region
of Fig.~\ref{fig:results}d.
This effect arises due to the capacitive coupling to the reservoir, 
as shown by the following argument. 
First, consider a specific setting:
the non-equilibrium situation when $N=0$, $z=0$, 
the top gate is grounded, $V_\text{t} = 0$, 
and the bottom gate voltage is
negative, $V_\text{b} < 0$.
That corresponds to $\delta V > 0$ and $\bar{V} < 0$,
i.e., the region where the negative $z_\text{eq}$ is
observed in Fig.~\ref{fig:results}d.
In this case, a charge $q_\text{r}$ 
is accumulated on the plate of the reservoir, 
and the charges accumulated on the
three plates r, t, b associated to the  dot
(see Fig.~\ref{fig:setup}c)
are $-q_\text{r}$, $-q_\text{r}$ and $2 q_\text{r}$, 
respectively.
This means that the Coulomb attraction between the 
plates of the bottom capacitor is four times stronger
than for the top capacitor, implying that
the dot is pulled toward the bottom gate.
In a more general case, when the gate voltages 
are not specified, but $\delta V > 0$ and $\bar{V} < 0$ still hold, 
we can say that a finite charge will accumulate on the 
plate facing the reservoir, and therefore the sum of the 
charges on the plates facing the top and bottom gates 
does not vanish. 
In turn, that sum determines the force acting on the dot, 
hence we conclude that that force is nonzero, 
and therefore the dot is displaced. 

It is expected that transport, e.g., electron-pair tunneling, 
through such a suspended quantum dot will be 
sensitive to the size of the jumps 
on the displacement stability diagram:
the larger the displacement jump, the lower 
the current flowing through the device. 
This effect is known as the Franck-Condon blockade
\cite{Braig,KochFC,Koch2006,Hwang_pairtunneling,Leturcq}, 
and arises for the following reason. 
In a transport situation, the system is voltage-tuned to
a point along one of the instability lines, e.g., the 0/2 boundary. 
Then, electrons can hop between the 
leads and the dot, and hence the occupation of the dot can 
fluctuate between 0 and 2.
As argued above, the ground-state displacement for the
two occupancies is different. 
As a consequence, the overlap between the 
resonator's quantum states corresponding to the 
two charge occupancies can become much 
smaller than unity; 
the larger the displacement jump at the selected point 
of the 0/2 boundary, 
the stronger the suppression of that overlap. 
In turn, that overlap controls the electron's tunnel rates between
the leads and the dot, and hence these tunnel rates and 
thereby the current through the device are also suppressed.
The quantitative characterization of these effects in 
the presence of 
mechanical nonlinearities, multiple mechanical modes,
finite temperature, and arbitrary electronic occupation is 
an important future theory task. 

\begin{figure*}
\centering
\includegraphics[width=\linewidth]{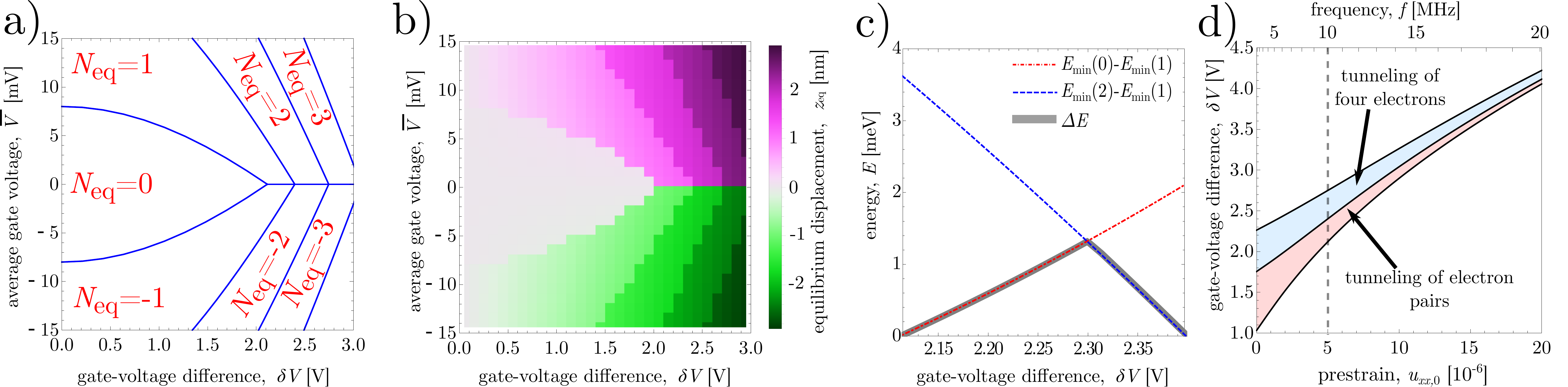}
\caption{
\textbf{
Results for a suspended metallic quantum dot.}
(a) Charge stability diagram. 
(b) Displacement stability diagram \cite{footnote}.
(c) Charge excitation gap $\Delta E$ (thick gray line) 
along the 1/-1 boundary shown in (a). 
Parameters: see caption of Fig.~\ref{fig:results}.
(d) Gate-voltage ranges of the 1/-1 and 2/-2 boundaries, 
as a function prestrain. 
The upper horizontal axis corresponds to the frequency $f$
associated to the resonator (see Methods). 
\label{fig:resultsmetallic}}
\end{figure*}

\emph{Results for a metallic suspended quantum dot.}
So far, we presented and discussed the results 
corresponding to an n-type semiconducting quantum dot, 
where the number of excess electrons is 
restricted to nonnegative integers, $N \geq 0$. 
However, depending on the specifics of the experimental setup, 
it can happen that the quantum dot is occupied by 
a large number $N_0$ of, say, conduction-band electrons,
when the top and bottom gate voltages are set to zero. 
In that case, which we refer to as a 
`metallic' dot\cite{Ojanen}, the number of these electrons can not only
be increased, but also 
decreased by tuning the system via the 
gate voltages. 
To characterize this scenario, in Fig.~\ref{fig:resultsmetallic}
we show the results corresponding to the 
metallic dot. 
To obtain the results in Fig.~\ref{fig:resultsmetallic}, 
we simply extended the energy minimization procedure
(described in Methods) to include
negative occupancies, $N \in \{-4, -3, \dots, 3, 4\}$. 

There are two major differences between the metallic 
(Fig.~\ref{fig:resultsmetallic})
and semiconducting
(Fig.~\ref{fig:results}) results.
(i) Naturally, the stability diagrams of the metallic dot
show perfect antisymmetry with respect to the
average gate voltage $\bar{V}$; this antisymmetry is absent
in the semiconducting case.
Note that it is expected that the $\bar{V} > 0$ 
part of the semiconducting and metallic stability 
diagrams are identical, and that expectation is confirmed by
comparing
Fig.~\ref{fig:results}c with 
Fig.~\ref{fig:resultsmetallic}a
and
Fig.~\ref{fig:results}d with
Fig.~\ref{fig:resultsmetallic}b.
(ii) The charge stability diagram of the metallic dot
(Fig.~\ref{fig:resultsmetallic}a)
shows boundaries of the type $N/-N$, along the
$\bar{V} = 0$ line. 
In principle, tuning the system to such a boundary
could imply transport via tunneling of electron packets
of size $2N$. 

\section{Discussion}

Steady advances in nanofabrication now allow the tailoring 
of electron-phonon interaction in suspended 
quantum dots\cite{Benyamini2014}. 
For graphene-based mechanical resonators, 
the tension and hence the nonlinearity 
can be tuned \emph{in situ} \cite{DeAlba2016} 
and the real-space mode shape can be visualized \cite{Davidovikj2016}. 
Suspended nanostructures with top and bottom gates
have been fabricated and studied in 
various experiments\cite{Allen2012,Leturcq,Weber_cntvibrations}.
In addition, quantum dots on suspended graphene ribbons 
have been created \cite{Allen2012},
and voltage-controlled charge-phonon coupling in such devices have
been demonstrated\cite{GangLuo}. 
Such devices bear the promise of 
combining few-electron transport with the outstanding 
mechanical
characteristics, e.g., high Q-factor and low mass density, of 
this material\cite{Eichler2011}.
Similar structures could be fabricated using members of
the recently discovered family of two-dimensional 
materials\cite{CastellanosGomez_adpreview}.
These developments suggest that  the fabrication 
of an engineered device as proposed here 
is within reach. 

A conceptually simple experiment to observe the charge stability
diagram in Fig.~\ref{fig:results}c could be based on charge sensing;
that is, $N_\text{eq}$
could be measured via 
the current flowing through a mesoscopic conductor that
is capacitively coupled to the suspended dot and therefore 
sensitive to $N_\text{eq}$.
Without the charge sensor, the boundaries of the diagram 
could also be mapped by sending a current through the 
suspended quantum dot, e.g., by 
utilizing the left and right reservoir depicted in Fig.~\ref{fig:setup}b
as a source and drain contact, respectively.
For a fixed value of the gate-voltage difference
below the $N_\text{eq} = 0,1,2$ triple point of Fig.~\ref{fig:results}c,
that is, for $\delta V < \delta V_2 \approx 2.6\, \text{V}$, 
the current $I(V_\text{sd},\bar{V})$
at finite bias voltage $V_\text{sd}$ would show
standard Coulomb-blockade features.
However, in the range $\delta V_2  < \delta V < \delta V_3$, 
the characteristics of electron-pair tunneling\cite{Koch2006,Hwang_pairtunneling}
discussed in the introduction 
are expected to appear when $I(V_\text{sd},\bar{V})$ is
measured.

Our results indicate that such a device would allow for
the exploration of transport via multi-electron tunneling as well. 
For example, when voltage-tuned to the 0/3 boundary
of Fig.~\ref{fig:results}c, 
current could be carried by the simultaneous tunneling
of three electrons. 
To our knowledge, it is an open challenge for both theory and
experiment to characterize such exotic scenarios. 

So far, our discussion focused on the case of zero temperature.
To observe a sharp 0/2 boundary 
of the charge stability diagram shown in 
Fig.~\ref{fig:results}c, 
the thermal energy scale $k_B T$ should be much lower
than the charge excitation gap $\Delta E$ along the 
0/2 boundary.
Along this boundary, the charge excitation gap is the
difference between the energy 
of the lowest-energy excited charge configuration, 
being the lower of 
$E_\text{min}(N=1)$ and 
$E_\text{min}(N=3)$,
and that of the twofold degenerate ground state
$E_\text{min}(N=0) =  E_\text{min}(N = 2)$.
Here, $E_\text{min}(N) = \min_{z} E(N,z)$. 
We plot the charge excitation gap in Fig.~\ref{fig:results}e, 
for the parameter set listed in the caption.
The plot shows $\Delta E$ (thick gray line) as the two gate voltages
are varied simultaneously such that we move along 
the 0/2 boundary.
For this example, the charge excitation gap is tuned between zero,
at the triple points, to a maximum of $\approx 2.2 \, \text{meV}$,
reached around the center 
of the considered gate-voltage range,
at $\delta V \approx 2.95 \, \text{V}$. 
Regarding the fact that temperatures of the order of 100 mK
corresponding to an energy scale of $k_B T \approx 10 \, \mu\text{eV}$
are available, reaching the above condition $k_B T \ll \Delta E$ 
seems experimentally feasible. 
Similar conclusion can be reached in the case of the metallic
dot, see Fig.~\ref{fig:resultsmetallic}c.

Figure \ref{fig:results}c proves the existence of the 0/2
and 0/3 boundaries in the charge stability diagram for  
a specific parameter set, in case of the semiconducting dot. 
We demonstrate the robustness of these boundaries with
respect to parameter variations by revealing how they 
change as the prestrain $u_{xx,0}$ of the ribbon is varied. 
In Fig.~\ref{fig:results}f, we plot the prestrain 
dependence of the gate-voltage-difference
coordinates 
characterizing the three triple points of Fig.~\ref{fig:results}c,
that is,  
$\delta V_2$, $\delta V_3$, $\delta V_4$.
Note that the prestrain value $u_{xx,0} = 5 \times 10^{-6}$
corresponds to Figs.~\ref{fig:results}c,d,e.
Figure \ref{fig:results}f indicates that 
the gate-voltage intervals 
$[\delta V_2,\delta V_3]$ and
$[\delta V_3,\delta V_4]$ of the 0/2 and 0/3 boundaries 
shrink as the prestrain is increased.
However, in the considered range of prestrain, 
the order of magnitude of the required 
gate-voltage difference remains to be a few volts, 
and the order of magnitude of the 
widths $\delta V_3 - \delta V_2$
and $\delta V_4 - \delta V_3$ 
remains to be a few hundred millivolts. 
The corresponding result in the case of the metallic dot
are shown in Fig.~\ref{fig:resultsmetallic}d.

\begin{figure}
\centering
\includegraphics[width=\columnwidth]{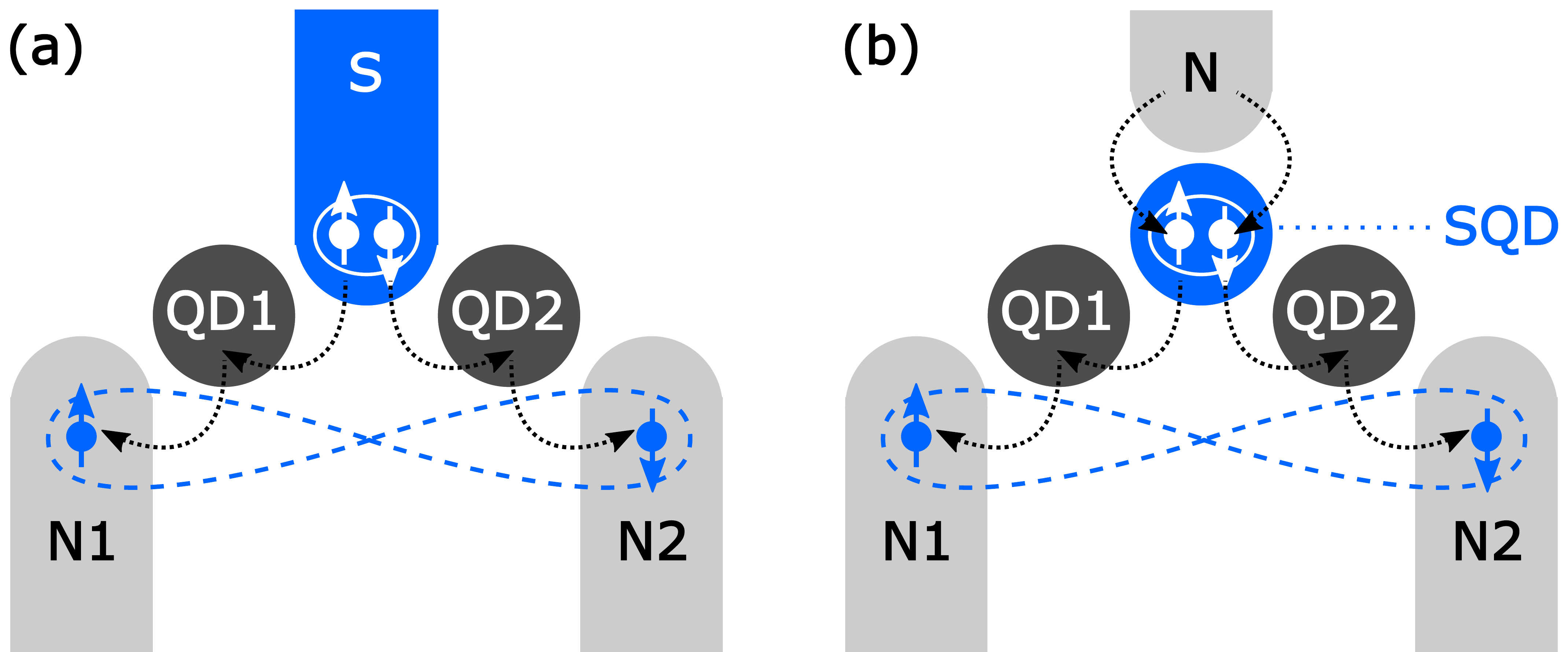}
\caption{
\textbf{
Mechanically assisted electron-electron attraction for 
creating a stream of spatially separated
spin-entangled electrons.}
(a) A Cooper-pair splitter circuit, based on the
controlled emission of Cooper pairs 
from a superconducting source electrode (S),
through two quantum dots (QD1 and QD2),
to two different normal drain electrodes (N1 and N2). 
(b) A similar functionality is offered if the 
superconducting electrode is replaced by 
a normal electrode (N) and a suspended quantum dot (SQD)
that is tuned to the 0/2 boundary of its charge stability
diagram.
\label{fig:cps}}
\end{figure}

As  pointed out above, 
it is expected that transport through a suspended 
quantum dot can take the form of sequential tunneling
of electron pairs, in a fashion reminiscent of 
certain electronic circuits containing
superconducting leads\cite{Jehl}.
This possibility naturally leads to the question: 
are there any electronic arrangements, where
the functionality of a superconducting lead can
be mimicked by a suspended quantum dot? 
In Fig.~\ref{fig:cps}, we present such a setup.
Fig.~\ref{fig:cps}a shows the sketch of a Cooper-pair splitter 
circuit\cite{Recher-andreev,Hofstetter,Herrmann},
where a superconducting
source electrode serves as a source of spin-singlet 
Cooper pairs, and the two quantum dots transfer
the electrons to the left and right normal leads such
that the entanglement of their spins is maintained.
Hence, this device supplies a stream of 
spatially separated, spin-entangled pairs of electrons.
It has been suggested that this functionality can 
be achieved even if the superconducting source
is replaced by a normal-metal electrode, 
and a third (conventional) quantum dot\cite{Saraga}. 
Here, we suggest to utilize the mechanical 
degree of freedom of the suspended 
dot for the same purpose, 
in the setup shown in Fig.~\ref{fig:cps}b.
The suspended quantum dot 
should be gate-voltage-tuned to the 0/2 boundary, 
which could guarantee that only spin-singlet 
pairs of electrons can 
tunnel from the source to the triple-dot
system, and thereby allow for the creation of a stream
of spatially separated spin-entangled electrons.

\section{Methods}

\emph{Anderson-Holstein model:
attractive electron-electron interaction and stability diagrams.}
In the Introduction, we discussed that an effectively attractive
electron-electron interaction can arise\cite{Anderson1975}
in the Anderson-Holstein
model depicted in Fig.~\ref{fig:setup}a.
Here, we summarize how that conclusion is reached,
and outline how the charge 
and displacement stability diagrams shown in Fig.~\ref{fig:results}a
and b are obtained.

As discussed in the Introduction, the 
energy of the Anderson-Holstein model is the sum of 
three contributions, $E_\text{AH} = 
E_\text{AH,m} + E_\text{AH,o} + E_\text{AH,int}$.
The mechanical energy is 
$E_\text{AH,m} = \frac 1 2 m \omega^2 z^2$,
the energy of the electrons occupying
the orbital is 
$E_\text{AH,o} = \epsilon N + \frac{U}{2}N(N-1)$, 
whereas the electron-phonon interaction energy is 
$E_\text{AH,int} = \lambda z N$. 

The energy of the Anderson-Holstein model can be 
rewritten\cite{Anderson1975,Koch2006,Ojanen}, using
the definitions $z_0 = \lambda / (m\omega^2)$,
$\epsilon_\text{eff} = \epsilon - \lambda^2/(2m\omega^2)$, and 
$U_\text{eff} = U- \lambda^2/(m\omega^2)$, 
as 
\begin{eqnarray}
E_\text{AH}= \frac 1 2 m \omega^2 (z+Nz_0)^2 + \epsilon_\text{eff} N
+ \frac 1 2 U_\text{eff} N(N-1).
\end{eqnarray}
This energy function is the same as that of a
system where the harmonic oscillator has an occupation-dependent
equilibrium position at $-N z_0$, 
and the electronic orbital is characterized
by the on-site energy $\epsilon_\text{eff}$ and 
the electron-electron interaction energy $U_\text{eff}$. 
As mentioned in the main text, 
$U_\text{eff}$ becomes negative, and hence can be interpreted
as an attractive electron-electron interaction, for sufficiently 
strong electron-phonon coupling strengths 
$\lambda > \omega\sqrt{ m U}$. 

The charge and displacement stability diagrams 
in Fig.~\ref{fig:results}a and b characterize the 
lowest-energy state of the electron-phonon system 
composed with 
a zero-temperature electron reservoir
with Fermi energy $\mu = 0$ (Fig.~\ref{fig:setup}a).
The occupation $N_\text{eq}$ 
and displacement $z_\text{eq}$ of the lowest-energy
states are plotted in Fig.~\ref{fig:results}a and b. 
These are analytical results, 
obtained by minimizing the energy function $E_\text{AH}(N,z)$
with respect to $N$ and $z$.

\emph{Charge and displacement 
stability diagrams of the suspended quantum dot.}
The stability diagrams  in 
Figs.~\ref{fig:results}c and d are
obtained by minimizing the total energy $E(N,z)$ of the suspended
quantum dot depicted in Fig.~\ref{fig:setup}b.
This energy 
$E = E_\text{m} + E_\text{em} + W$ 
includes a purely mechanical ($E_\text{m}$), 
an electromechanical ($E_\text{em}$), 
and a purely electronic ($W$) contribution.
Here we describe how these contributions are estimated, 
and how the energy function is used to obtain the stability diagrams.

First, we describe the purely mechanical contribution, 
associated to the deformation of the graphene ribbon. 
We assume that most of the excess charge 
of the quantum dot is localized in a narrow region
around the center of the suspended 
part of the ribbon. 
Then, the gate-induced forces stretch the ribbon in the way shown in 
Fig.~\ref{fig:setup}b, and the deformation-induced
elongation of the ribbon is 
assumed to be homogeneous for simplicity. 
The mechanical energy arising from this stretching 
deformation can be expressed as a function of the
ribbon's parameters and the displacement $z$ of the dot:
\begin{eqnarray}
\label{eq:emech}
E_\text{m}(z) = \frac 1 2 Y W L_u u^2_{xx}(z).
\end{eqnarray}
Here, $Y = 340 \, \text{N}/\text{m}$ 
is the Young modulus of graphene\cite{Lee2008},
and $u_{xx}(z) = (L(z) - L_u)/L_u$ is the
relative elongation (strain) of the ribbon. 
Simple geometrical considerations imply 
\begin{equation}
\label{eq:uxx}
u_{xx}(z) = \sqrt{(1+u_{xx,0})^2 + 4(z/L_u)^2} -1.
\end{equation}
Substituting this into Eq.~\eqref{eq:emech} and
expanding the latter up to fourth order in $z$, 
we find, up to a constant, 
\begin{eqnarray}
\label{eq:energym}
E_\text{m}(z) \approx 
\alpha_2 z^2 + \alpha_4 z^4,
\end{eqnarray}
where 
$\alpha_2 \approx 2 Y u_{xx,0} W / L_0$ and
$\alpha_4 \approx 2Y W / L_0^3$;
the latter two expressions are accurate up to leading order in 
the small prestrain $u_{xx,0} \ll 1$.

The electromechanical contribution to the energy is
associated to the effective capacitors\cite{Nazarov_book} shown
in Fig.~\ref{fig:setup}c: 
\begin{eqnarray}
\label{eq:electromechanical}
E_\text{em}(N,z) = 
q_{\rm r}^2/2C+q_{\rm t}^2/2C_{\rm t}(z)+q_{\rm b}^2/2C_{\rm b}(z),
\end{eqnarray}
where $q_\text{r,t,b}$ are the charges accumulated on the
reservoir, top gate, and bottom gate, respectively. 
The relation of these charges to the dot occupancy $N$ and
displacement $z$ can 
be established using that
(i) the  quantized dot charge $-|e| N$ can 
be expressed as 
$ - |e| N= 
- q_\text{r} - q_\text{t} - q_\text{b} 
$,
(ii) the top gate voltage is the sum of the voltages
dropping on the reservoir and top capacitors, 
$V_\text{t} = q_\text{t} / C_\text{t}(z) - q_\text{r}/C$,
(iii) analogously for the bottom gate voltage, 
$V_\text{b} = q_\text{b} / C_\text{b}(z) - q_\text{r}/C$.
This linear set of three equations for $q_\text{r,t,b}$ is solved, 
and the solutions are inserted into Eq.~\ref{eq:electromechanical},
to obtain the explicit dependence of the 
electromechanical energy $E_\text{em}$ on $N$ and $z$. 

The third, last contribution to the total energy of the system is the
work done by the voltage sources\cite{Nazarov_book}
$W(N,z) = - q_\text{t} V_\text{t} - q_\text{b} V_\text{b}$. 

To find the equilibrium occupation $N_\text{eq}$ and
displacement $z_\text{eq}$ of the suspended quantum dot, 
we minimize the energy $E(N,z)$ as a function of displacement $z$ 
and occupation number $N$, using the following 
procedure. 
We focus on the case of small displacements $z\ll d$, 
hence we Taylor-expand the electromechanical $E_\text{m}$ and 
the electronic $W$ terms up to second order 
in the variable $z$ around zero. 
By this expansion, the total energy function becomes 
a fourth-order polynomial of $z$, which we can  minimize 
numerically with respect to $z$ for the different occupation 
numbers $N\in\{0,1,2,3,4\}$, yielding the minimum value
$E_\text{min}(N) = \min_z E(N,z)$ of the energy and
the corresponding displacement 
$z_\text{min}(N)$, assuming dot occupancy $N$.
The equilibrium occupation $N_\text{eq}$ at zero temperature is 
then found my minimizing the energy $E_\text{min}(N)$,
also yielding the equilibrium displacement 
$z_\text{eq} = z_\text{min}(N_\text{eq})$.
Repeating this procedure for various values of the gate voltages, 
we obtain the charge and displacement stability diagrams 
shown in Fig.~\ref{fig:results}c,d, respectively.
The same procedure is followed to obtain the
results for the metallic dot, Fig.~\ref{fig:resultsmetallic}a,b, 
with the generalization that we allow for negative occupation
numbers as well, $N \in \{-4, -3, \dots, 3, 4\}$. 

\emph{Relation between the Anderson-Holstein model and 
the suspended quantum dot model.} 
Here, we establish 
the relation between the Anderson-Holstein
model and the 
semiconducting suspended quantum dot model. 
Focusing on the regime of large gate-voltage differences, 
e.g., to the vicinity of the 0/2 boundary where 
electron pair tunneling is expected, we show that
the energy of the suspended quantum dot model
incorporates two terms that are 
absent in the energy of the Anderson-Holstein model.
 
As mentioned above, the energy of the suspended 
quantum dot $E(N,z)$ in the small displacement regime $z\ll d$ 
can be approximated by its Taylor expansion in the variable $z$. 
This yields
\begin{eqnarray} \label{TaylorE}
E(N,z)&\approx&\alpha_2z^2+\frac{ e^2 N^2}{6C}
-\frac{2}{3}|e|N  \bar{V}
-\frac{2}{3} |e| N \delta V \frac z d \nonumber
\\
&-&
\frac{1}{3}C \delta V^2
	\left(\frac{z}{d}\right)^2
-\frac{2}{3}C\delta V \bar{V} \frac z d 
+\alpha_4z^4,
\end{eqnarray}
where  terms independent of $N$ and $z$ are neglected. 
In Eq.~\eqref{TaylorE} we drop further 
second- and higher-order terms in $z$, 
which, in the vicinity of the 0/2 boundary where
$|\bar{V}|, |e| / C \ll \delta V$ holds,
 are much smaller than $C\delta V^2(z/d)^2$. 

We now compare the energy 
$E$ of the quantum dot model in Eq.~\eqref{TaylorE}, 
with the energy $E_\textrm{AH}$ of the Anderson-Holstein model.
We claim that the first five terms of $E$ 
correspond to the four terms of $E_\text{AH}$. 
That is, the former  is obtained from the latter
via making the substitutions
\begin{eqnarray}
U&\mapsto& \frac{e^2}{3C} ,\\
\label{eq:ahsqd_frequency}
\frac{1}{2}m\omega^2 &\mapsto& \alpha_2-\frac{1}{3}C\delta V^2
\frac{1}{d^2}, \\
\label{eq:onsite}
\epsilon &\mapsto& -\frac{2|e|\bar{V}}{3} + \frac{e^2}{6C}, \\
\label{eq:coupling}
\lambda &\mapsto& -\frac{2}{3}|e| \delta V \frac 1 d.
\end{eqnarray}
These results confirm the expectations that
the average gate voltage $\bar{V}$ controls the on-site 
energy $\epsilon$ [Eq.~\eqref{eq:onsite}], 
and the gate-voltage difference 
$\delta V$ controls the electron-phonon coupling strength $\lambda$
[Eq.~\eqref{eq:coupling}]. 
The second term on the right-hand side of 
Eq.~\eqref{eq:ahsqd_frequency} shows that
the oscillator eigenfrequency of the Anderson-Holstein model
corresponds to 
a combination of a mechanical and an electronic contribution
in the suspended-dot model. 
The last two terms in the energy Eq.~(\ref{TaylorE}) of the 
suspended quantum dot, i.e., 
the charge-independent term proportional to the displacement
and the quartic potential, 
do not appear in the Anderson-Holstein model.

\emph{Resonator frequency.}
The frequency $f$ associated to the resonator in
Fig.~\ref{fig:results}f is defined from the quadratic 
term of the mechanical energy $E_\text{m}$ 
in Eq.~\ref{eq:energym} via 
$\alpha_2 z^2 = \frac 1 2 m (2\pi f)^2 z^2$.
Here, $m = W L_0 \rho$,
which approximates the mass $W L_u \rho$ of the 
suspended part of the graphene ribbon, 
with the surface mass density of graphene
being $\rho =
7.61 \times 10^{-7} \, \text{kg}/\text{m}^2$.
From these, the frequency is expressed as
\begin{equation}
f=\frac{1}{\pi L_0 } \sqrt{\frac{Y u_{xx,0}}{\rho}}.
\end{equation}

\emph{Extrinsic forces can be much stronger than intrinsic ones.}
The force acting on a singly-occupied 
suspended quantum dot due to
the extrinsic electron-phonon interaction, i.e., due to
the electric field induced by the top and bottom gates, 
is estimated as
$F_\text{e} = e(V_\text{t}-V_\text{b}) / 2 d$.
Inserting the characteristic values 
$V_\text{t} - V_\text{b} = 1 \, \text{V}$ and
$d = 150 \, \text{nm}$, we find $F_\text{e} = 3.3\, \text{meV}/\text{nm}$. 
On the other hand, in equilibrium, the force acting on the quantum dot 
due to the dominant intrinsic electron-phonon
coupling, that is, the deformation potential mechanism, is
$F_\text{i} = 
\left\{\partial_z \left[ \Xi u_{xx} (z) \right]\right\}_{z=z_\text{eq}}
\approx 4 \Xi z_\text{eq}/L_0^2$.
Here, 
$\Xi = 30 \, \text{eV}$ is the in-plane deformation potential 
of graphene\cite{Suzuura2002,Droth2016},
we used
Eq.~\ref{eq:uxx}, and we present the leading-order result 
in the small quantities $z_\text{eq}/L_0$ and $u_{xx,0}$.
For the characteristic values of $z_\text{eq} = 1 \, \text{nm}$ and
$L_0 = 1.5\,  \mu\text{m}$
(see Fig.~\ref{fig:results}d), we
find $F_\text{i} \approx 0.05 \, \text{meV}/\text{nm}$, 
which indeed fulfills $F_\text{i} \ll F_\text{e}$.
In the suspended quantum dot model, we have neglected the 
energy contribution of the intrinsic electron-phonon interaction, 
and that simplification is justified by these quantitative
estimates.

\section{Conclusions}

In conclusion, we suggested a way to engineer an
electromechanical system that exhibits effective
electron-electron attraction. 
Our study, focused  on the example of
a suspended quantum dot in a graphene nanoribbon, 
supports the experimental feasibility of observing the remarkable
but so far elusive
equilibrium and transport phenomena implied by the
attractive nature of the  interaction.
Furthermore, our work suggests 
the possibility that certain functionalities
of superconducting nanostructures can be achieved 
by substituting the superconducting
elements with appropriately
assembled electromechanical systems. 
These results raise interesting questions 
regarding, e.g., 
the feasibility of realizing an electron-based
quantum simulator of the attractive Hubbard model, 
or the design of artificial superconductors
based on engineered electromechanical systems as
building blocks.

\acknowledgments

We thank
G. Rastelli, P. Simon, and S. Weiss for useful discussions.
GSz acknowledges financial support of the
National Research, Development and Innovation Office of 
Hungary via the National Quantum Technologies Program 
NKP-2017-00001 and the OTKA Grant 108676.
AP acknowledges funding from the EU Marie Curie Career Integration
Grant CIG-293834 (CarbonQubits), the OTKA Grants
105149 and
PD 100373,
and the EU ERC Starting Grant CooPairEnt 258789.
MD acknowledges funding from the 
Deutsche Forschungsgemeinschaft 
(DFG, German Research Foundation) within Project 
No.~\mbox{317796071}.


\begin{thebibliography}{}
\bibitem{footnote}
The pixel structure in Figs.~\ref{fig:results}d and 
\ref{fig:resultsmetallic}b is due to
data being obtained on a $\delta V$-$\bar{V}$ grid
of size $36 \times 30$.
\end{thebibliography}
\end{document}